\documentclass{article}
\usepackage{spconf,amsmath,graphicx,amsfonts,cite,tabularx,multirow,xcolor,url,tipa,breakurl}
\usepackage[breaklinks]{hyperref}

\newcolumntype{D}{>{\centering\arraybackslash}m{7.5ex}}
\newcolumntype{E}{>{\centering\arraybackslash}m{7ex}}

\title{G2G: TTS-DRIVEN PRONUNCIATION LEARNING FOR GRAPHEMIC HYBRID ASR}
\name{Duc Le, Thilo Koehler, Christian Fuegen, Michael L. Seltzer}
\address{Facebook AI\\
	{\small \texttt{\{duchoangle,tkoehler,fuegen,mikeseltzer\}@fb.com}}}

\begin{document}
%
\maketitle
\begin{abstract}
Grapheme-based acoustic modeling has recently been shown to outperform phoneme-based approaches in both hybrid and end-to-end automatic speech recognition (ASR), even on non-phonemic languages like English. However, graphemic ASR still has problems with low-frequency words that do not follow the standard spelling conventions seen in training, such as entity names. In this work, we present a novel method to train a statistical grapheme-to-grapheme (G2G) model on text-to-speech data that can rewrite an arbitrary character sequence into more phonetically consistent forms. We show that using G2G to provide alternative pronunciations during decoding reduces Word Error Rate by 3\% to 11\% relative over a strong graphemic baseline and bridges the gap on rare name recognition with an equivalent phonetic setup. Unlike many previously proposed methods, our method does not require any change to the acoustic model training procedure. This work reaffirms the efficacy of grapheme-based modeling and shows that specialized linguistic knowledge, when available, can be leveraged to improve graphemic ASR.
\end{abstract}
\section{INTRODUCTION}
\label{sec:intro}

There is a growing trend in the automatic speech recognition (ASR) community to use graphemes directly as the output units for acoustic modeling instead of phonemes \cite{gales2015unicode,chan2015listen,DBLP:conf/interspeech/SakSRB15,Soltau17,sainath2017no,zeyer2018improved,Chiu18,Audhkhasi18Word,Li2018WordCTC,Ueno2018WordCTC,He2019RNNT,Irie2019output,Le2019Kulfi,Wang20KulfiTransformer}. Grapheme-based modeling, which does not rely on any specialized linguistic information, has been shown to outperform phoneme-based modeling in both end-to-end \cite{sainath2017no,Irie2019output} and traditional hybrid ASR \cite{gales2015unicode,Le2019Kulfi}, even on non-phonemic languages with poor grapheme-phoneme relationship like English \cite{Le2019Kulfi}. Despite achieving better overall recognition accuracy, grapheme-based modeling still has problems with rare words that are pronounced differently than how they are spelled or do not conform to the standard spelling conventions, most notably proper nouns. These problems are typically addressed in phoneme-based approaches by having linguists manually correcting the pronunciations and training a grapheme-to-phoneme (G2P) model to generalize to previously unseen words. Thanks to this extra linguistic information, phonetic ASR may perform better than grapheme-based approaches in long-tail name recognition. It is therefore appealing to leverage linguistic knowledge in the same way to improve graphemic ASR's performance on rare entity names.

In this work, we propose a novel method to distill linguistic knowledge into graphemic ASR by automatically learning pronunciations at the grapheme level on artificial audio generated with text-to-speech (TTS). The outcome of this method is a statistical grapheme-to-grapheme (G2G) model that can transform a character sequence into homophones with more conventional spelling, such as rewriting ``Kaity" to ``Katie." We show that using G2G to generate alternative pronunciations during decoding results in \textbf{3\%} to \textbf{11\%} relative Word Error Rate (WER) improvement for graphemic models on utterances containing entity names. With G2G, our graphemic ASR system is able to bridge the gap with equivalent phonetic baselines on rare name recognition while achieving better overall WER. Our work addresses a long-standing weakness of grapheme-based models and contributes a novel way to combine the strengths of graphemic and phonetic ASR.

\section{RELATED WORK}
\label{sec:related_work}

Grapheme-based modeling with word pieces has become the standard approach for end-to-end ASR, outperforming both context-independent phonemes and graphemes (e.g., \cite{sainath2017no,zeyer2018improved,Chiu18,He2019RNNT,Irie2019output}). More recently, we showed that context- and position-dependent graphemes (i.e., \textit{chenones}) are also extremely effective for hybrid ASR, significantly outperforming senones \cite{Le2019Kulfi} and achieving state-of-the-art results on Librispeech \cite{Wang20KulfiTransformer}. In this work, we continue to improve graphemic hybrid ASR performance on name recognition.

Existing solutions for handling proper nouns and rare words for graphemic ASR have mostly been done in the context of end-to-end systems and typically involve a TTS component to generate additional synthetic data for AM training \cite{He2019RNNT,Peyser2019,Zhao2019,Rosenberg19ASR+TTS}. While the resulting improvement is promising, this approach requires careful balancing between real and synthetic data to prevent overfitting to TTS, large amount of synthesized data (relative to real data), as well as a highly diversified artificial speaker pool, all of which significantly complicate the AM training process. A different class of approach involves training a neural spelling correction model on TTS data to fix errors made by Listen, Attend, and Spell (LAS) ASR models \cite{Guo2019}. While this technique does not complicate AM training, it does not explicitly address the proper noun recognition problem and the WER gain they achieved on Librispeech was limited. Other methods that do not rely on TTS include leveraging phonetic information to build better word piece inventories \cite{Xu2019} and fuzzing the training data with phonetically similar words \cite{Alon2019,Zhao2019}.

Our proposed approach differs from previous work in three ways. First, it also leverages TTS but instead produces alternative pronunciations that are ingested on-the-fly during decoding, thus does not require any change to AM training and preserves the system's simplicity. Second, our method is geared toward conventional hybrid ASR whereas previous work mostly focused on end-to-end. Third, our work directly resolves the mismatch between graphemes and acoustic models, the fundamental cause of name recognition errors which previous work only addressed indirectly.

\section{DATA}
\label{sec:data}

\subsection{Audio Data}
\label{ssec:audio_data}

Our training data contains a mixture of two in-house hand-transcribed anonymized datasets with no personally identifiable information (PII). The first dataset consists of 15.7M utterances (12.5K hours) in the voice assistant domain recorded via mobile devices by 20K crowd-sourced workers. Each utterance is distorted twice using simulated reverberation and randomly sampled additive background noise extracted from public Facebook videos. The second dataset comprises 1.2M voice commands (1K hours) sampled from the production traffic of Facebook Portal\footnote{Portal is a video-enabled smart device that supports calling (e.g., ``hey Portal, call Alex") and other types of voice queries (e.g., ``hey Portal, what's the weather in Menlo Park?").} after the ``hey Portal" wakeword is triggered. To further de-identify the user, utterances from this dataset are morphed when researchers access them (the audio is not morphed during training). We use speed perturbation \cite{Ko2015AudioAF} to create two additional copies of each utterance at 0.9 and 1.1 times the original speed. Finally, we distort each copy with additive noise sampled from the same source described previously, resulting in six copies in total (\{0.9, 1.0, 1.1 speed\} $\times$ \{no noise, with noise\}). The total amount of data after distortion is 38.6M utterances (31K hours).

Our evaluation data consists of 54.7K hand-transcribed anonymized utterances from volunteer participants in Portal's in-house dogfooding program, which consists of employee households that have agreed to have their Portal voice activity reviewed and tested. Every utterance in this set has an associated contact list, which we use for on-the-fly personalization for calling queries. We further split this evaluation set into three subsets. Firstly, \texttt{name-prod} comprises 11.4K utterances with 3.9K unique entity names from the personalized contact list. Secondly, \texttt{name-rare} is similar to \texttt{name-prod}, but with 800 utterances and 700 unique entity names; it contains more diverse names than those typically observed in traffic and is designed to stress-test our ASR system on name recognition. Lastly, \texttt{non-name} consists of 42.5K utterances (10.6K unique transcripts) that do not have any entity name from the associated contact list. Note that utterances from this subset may still contain other types of entities not included in the personalization data, such as city names, song names, and artist names.

\subsection{Pronunciation Data}
\label{ssec:name_pron_data}

Our phonetic ASR system makes use of an in-house pronunciation corpus consisting of 1.1M unique written--pronunciation pairs, 75\% of which are name pronunciations, where the written form of a word is mapped to an X-SAMPA pronunciation \cite{Wells1995ComputercodingTI}. This dataset is used for G2P training and gives considerable advantage to phonetic ASR baselines in terms of name recognition since it has much wider name coverage than those included in the acoustic training data. The aim of this work is to distill the phonetic information from this pronunciation corpus into our graphemic ASR system.

\section{G2G TRAINING METHODS}
\label{sec:pron_learning}

\begin{figure}[tb]
  \centering
  \includegraphics[width=\columnwidth]{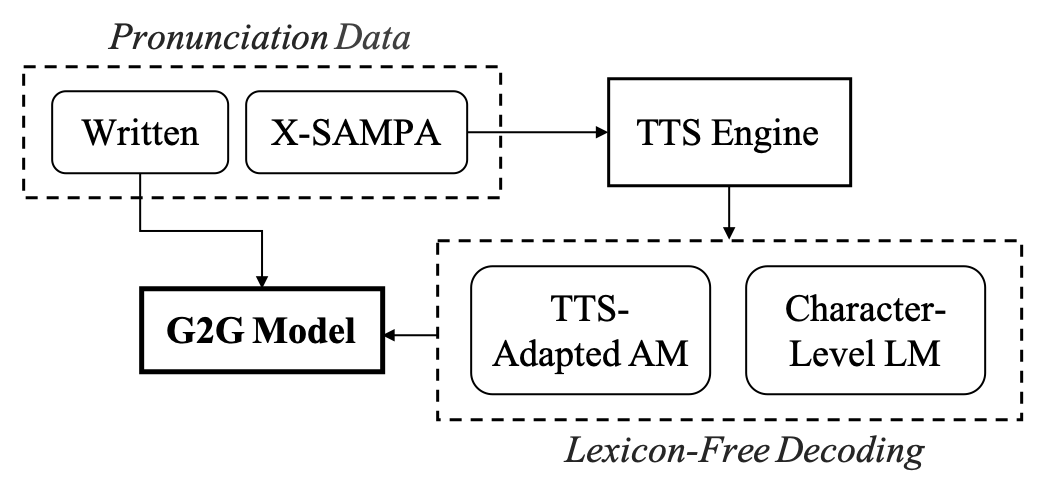}
  \caption{Overview of TTS-based G2G training pipeline.}
  \label{fig:diagram}
\end{figure}

In standard grapheme-based hybrid ASR, the G2P-equivalent operation is trivial; the pronunciation of a word (e.g., ``blue") is simply its decomposed grapheme sequence (e.g., ``b l u e"). However, this method may fail for long-tail words whose graphemes do not accurately reflect how they are pronounced. Having a G2P-like model for graphemes (i.e., \textit{G2G}), is therefore desirable to handle such cases. We hereby propose two training methods for G2G based on TTS and homophones.

\textbf{G2G-TTS}: Figure \ref{fig:diagram} summarizes our proposed approach to train a TTS-based G2G model, which consists of three main steps. First, we feed the X-SAMPA string for each entry in our pronunciation corpus through a TTS engine to obtain the corresponding vocalization. Note that we intentionally operate on X-SAMPA directly instead of the written form to bypass the reliance on TTS's internal G2P. Second, we do lexicon-free decoding (LFD) \cite{Likhomanenko2019} on these artificial utterances using a specialized character-level ASR system to generate grapheme sequences that accurately reflect their pronunciations while adhering to the English spelling convention. We will provide more detail about LFD in Section \ref{ssec:g2g_training}. Finally, we train a statistical G2G model to map the original written form of the pronunciation data (e.g., ``Kaity") to its LFD output (e.g., ``Katie"). This is in essence a sequence-to-sequence modeling problem with many viable approaches; however, we will employ the same method as G2P model training to ensure fair comparison with phonetic baselines.

\textbf{G2G-HOM}: the reliance on TTS is a limitation of the previous approach. One way to circumvent TTS is to gather clusters of homophones from our phonetic lexicon (e.g., \textit{Michael}, \textit{Mikall}, \textit{Mykol}) and train a model to map each word in the cluster to the cluster ``root" (e.g., \textit{Michael}). A cluster ``root" is defined to be the word that most closely adheres to the English spelling convention; the exact metric for this determination will be discussed in Section \ref{ssec:g2g_training}. Note that this shares some similarity with the ``transcript fuzzing" method utilized in \cite{Zhao2019}; however, unlike their approach which aims to increase variation during training by introducing more proper nouns, our approach aims to decrease variation during decoding.

\section{EXPERIMENTAL SETUP}
\label{sec:experiments}

\subsection{Baseline ASR Systems}
\label{ssec:baselines}

Our baselines are hybrid ASR systems with context- and position-dependent phonemes/graphemes similar to our previous work \cite{Le2019Kulfi}. For the phonetic setup, we train a joint-sequence G2P model \cite{bisani2008joint} on the pronunciation corpus described in Section \ref{ssec:name_pron_data}, using an in-house implementation of Phonetisaurus \cite{novak-etal-2012-wfst}. This model is used to generate pronunciations for both the training and decoding lexicons. For the graphemic setup, we preserve all casing information and use \texttt{unidecode} to normalize Latin characters. Excluding position-dependent variants, we have 47 phones and 56 graphemes, on which we train tri-context decision trees with 7K senones and 5K chenones.

The AM training procedure is identical across both phonetic and graphemic setups. We employ a multi-layer unidirectional Long Short-Term Memory RNN (LSTM) with five hidden layers and 800 units per layer (approximately 35M parameters). The input features are globally normalized 80-dimensional log Mel-filterbank extracted with 25ms FFT windows and 10ms frame shift; each input frame is stacked with seven right neighboring frames. The first LSTM hidden layer subsamples the output by a factor of three \cite{peddinti2018low} and the target labels are delayed by 10 frames (100ms). We first train the model using Cross Entropy (CE) for 15 epochs, followed by Lattice-Free Maximum Mutual Information (LF-MMI) \cite{povey2016purely} for 8 epochs. We use Adam optimizer \cite{kingma2014adam}, 0.1 dropout, Block-wise Model-Update Filtering (BMUF) \cite{Chen16BMUF}, and 64 GPUs for all AM training experiments.

Evaluation is done using our in-house one-pass dynamic decoder with n-gram language model (LM). We use a 4-gram class-based background LM with 135K vocabulary and 23M n-grams. During decoding, the \texttt{@name} class tag is expanded on-the-fly using the personalized contact list to allow for contextual biasing, and name pronunciations are obtained the same way as was done during training. For phonetic baselines, the number of pronunciation variants per name generated via G2P is a tunable hyperparameter (defaults to 2). For graphemic baselines, we add the lower-cased form of the name as an alternative pronunciation in addition to its original written form, producing up to two variants.

\subsection{G2G Training and Integration}
\label{ssec:g2g_training}

\begin{table}[t]
\centering
\begin{tabular}{ l l }
	Written & \texttt{interesting} \\
	LM & \texttt{i\_B n t e r e s t i n g\_E} \\
	Lexicon & \texttt{i\_WB n t e r e s t i n g\_WB} \\
\end{tabular}
\caption{Example character-level language model and lexicon entries in our lexicon-free decoding setup.}
\label{table:char_level_lm}
\end{table}

We follow the G2G training procedure outlined in Section \ref{sec:pron_learning}.

\textbf{G2G-TTS}: our TTS engine is an in-house single-voice system made up of a text-processing frontend, prosody models, acoustic models, and a neural vocoder, capable of producing high fidelity speech. For LFD, we employ a 10-gram character-level LM for decoding. This LM is trained on single words obtained from the training transcripts, which captures the spelling convention of English. To preserve position dependency, we augment the characters with additional position tags: \texttt{B} (beginning of word), \texttt{E} (end of word), and \texttt{S} (singleton). These position tags allow us to map each grapheme to the correct AM output unit; for example, the lexicon entry for \texttt{i\_B}, \texttt{i\_E}, and \texttt{i\_S} is \texttt{i\_WB} (\texttt{WB} stands for ``word boundary"), whereas the lexicon entry for \texttt{i} is \texttt{i} (see Table \ref{table:char_level_lm} for an example). The AM used for LFD is a Latency-Controlled Bidirectional LSTM (LC-BLSTM) \cite{Zhang16LCBLSTM} with five hidden layers and 800 units per direction per layer. This AM uses the same decision tree as our graphemic baseline; it is first trained on regular training data, then adapted with LF-MMI on TTS audio generated from the training transcripts. After adaptation, the Character Error Rate (CER) on a held-out TTS development set reduces from 1.64\% to 0.16\%, which denotes almost perfect recognition accuracy. The highly accurate AM coupled with the character-level LM ensures that the LFD output both captures the acoustic properties of the TTS data and closely adheres to the conventional English spelling.

\textbf{G2G-HOM}: our homophone-based G2G training data are 133.5K homophone clusters with at least two members, totaling 419.6K word pairs. We designate the cluster root to be the member with the highest normalized score obtained from the 10-gram character-level LM used for LFD. We hypothesize that high LM scores mean more canonical English spelling.

We apply the same G2P training method to train a joint-sequence G2G model. The resulting G2G and G2P models are therefore directly comparable; they use the same source data, the same underlying model, and the same training recipe. The output of G2G for personalized contact lists can either be used directly for decoding or mixed with the default graphemic pronunciations.

\section{RESULTS AND DISCUSSION}
\label{sec:results}

\begin{table}[t]
\centering
\begin{tabular}{c c || D | D D}
     \textbf{N} & \textbf{Dataset} & \textbf{G2P} & \textbf{G2G-HOM} & \textbf{G2G-TTS} \\
     \hline
     \hline
     \multirow{3}{*}{2} & \texttt{name-prod} & 9.6 & \multicolumn{2}{c}{9.7$^\dagger$} \\
     & \texttt{name-rare} & 9.6 & \multicolumn{2}{c}{10.3$^\dagger$} \\
     & \texttt{non-name} & 11.9 & \multicolumn{2}{c}{11.5$^\dagger$} \\
     \hline
     \multirow{3}{*}{3} & \texttt{name-prod} & 9.6 & 9.6 & 9.5 \\
     & \texttt{name-rare} & 9.1 & 10.4 & 9.7 \\
     & \texttt{non-name} & 11.9 & 11.5 & 11.5 \\
     \hline
     \multirow{3}{*}{4} & \texttt{name-prod} & 9.6 & 9.5 & 9.4 \\
     & \texttt{name-rare} & 9.0 & 10.2 & 9.3 \\
     & \texttt{non-name} & 11.9 & 11.5 & 11.5 \\
     \hline
     \multirow{3}{*}{5} & \texttt{name-prod} & \textbf{9.7}$^*$ & 9.5 & \textbf{9.4} \\
     & \texttt{name-rare} & \textbf{8.9}$^*$ & 10.0 & \textbf{9.2} \\
     & \texttt{non-name} & \textbf{11.9}$^*$ & 11.5 & \textbf{11.5} \\
\end{tabular}
\\
\vspace{1em}
{$*$: phonetic baseline}\space\space\space\space{$\dagger$: graphemic baseline without G2G}\\
{\textit{N}: maximum number of pronunciation variants}\\
\caption{WER comparison between phonetic (G2P) and graphemic (G2G-HOM, G2G-TTS) ASR systems.}
\label{table:asr_result}
\end{table}

\begin{table}[t]
\centering
\begin{tabular}{ l l }
	\textbf{Input} & \textbf{G2G-TTS Output} \\
	\hline
	\hline
	Kaity & \texttt{K\_WB a t i e\_WB} \\
	Ly & \texttt{L\_WB e e\_WB} \\
	Coce & \texttt{K\_WB o c h e\_WB} \\
	Sera & \texttt{S\_WB a r a h\_WB} \\
	Qifei & \texttt{C\_WB h i e f e\_WB} \\
	Liesl & \texttt{L\_WB e i s e l\_WB} \\
	\hline
	quake & \texttt{q\_WB u a k e\_WB} \\
	prosciutto & \texttt{p\_WB r o s h u t o\_WB} \\
	phoneme & \texttt{p\_WB h o n e e m\_WB} \\
	ASCII & \texttt{a\_WB s k y\_WB} \\
	\hline
\end{tabular}
\caption{Example G2G-TTS output for names (top half) and regular words (bottom half).}
\label{table:g2g_output}
\end{table}

As shown in Table \ref{table:asr_result}, the graphemic ASR baseline performs better by 3\% relative on \texttt{non-name} and similar on \texttt{name-prod} compared to the phonetic baseline, but underperforms significantly on \texttt{name-rare} by 16\% relative. This large gap on \texttt{name-rare} confirms the limitation of default graphemic pronunciations when dealing with unconventional names. For example, \textit{Kaity}, \textit{Ly}, \textit{Coce}, \textit{Sera}, \textit{Qifei}, and \textit{Liesl} are correctly recognized in the phonetic baseline but misrecognized in the graphemic baseline as \textit{Katie}, \textit{Lee}, \textit{Choquel}, \textit{Sarah}, \textit{Shi}, and \textit{Lisa}, respectively.

We next analyze the impact of G2G on ASR. We notice that the stand-alone G2G is good at handling the ``edge cases" while the default graphemic pronunciations are more appropriate for common names, thus we report results from combining the two\footnote{G2G-only results are consistently worse than the baseline numbers.}. Both G2G-HOM and G2G-TTS improve over the baseline, with the latter giving the best results, yielding an improvement of 3\% and 11\% on \texttt{name-prod} and \texttt{name-rare}, respectively. With this setup, the gap between phonetic and graphemic ASR on \texttt{name-rare} closes from 16\% to 3\% relative, while the latter performs better on the remaining test sets. We hypothesize that G2G-HOM underperforms because its output may not be fully compatible with the AM decision tree, whereas G2G-TTS guarantees this property through LFD. Further work is required to train a high-quality G2G system without relying on TTS.

We provide examples of G2G-TTS output in Table \ref{table:g2g_output} to gain more understanding of how the model works and why it helps ASR. The top half of the table shows the G2G output for the misrecognized names mentioned previously, all of which have been fixed with G2G. The model is able to rewrite these names into homophone variants with more conventional spellings. Notably, the model can tell that the ``Q" in \textit{Qifei} is pronounced as ``Ch;" this is a common source of errors for Chinese names in the baseline system. G2G also gives reasonable output for regular words as seen in the bottom half of the table, including spoken abbreviations like \textit{ASCII}. Given these examples, it will be promising to leverage G2G to improve AM training. Moreover, it will also be interesting to study the impact of G2G on end-to-end ASR systems based on word pieces or context-independent graphemes.

\section{CONCLUSION AND FUTURE WORK}
\label{sec:conclusion}

In this paper, we proposed a novel method for training a G2G model from TTS data that can rewrite words into homophone variants with more conventional spelling. We show that G2G output can be leveraged during decoding to significantly improve graphemic ASR's proper noun recognition accuracy and bridge the gap with phoneme-based systems on rare names. Future work will focus on further improving G2G, possibly by using neural sequence-to-sequence models, as well as applying G2G to AM training and end-to-end ASR.

\section{ACKNOWLEDGMENT}
\label{sec:acknowledgment}

We'd like to thank linguists Antony D'Avirro and Dat Vo for their help in formulating and curating data for the homophone-based G2G training approach.

\newpage
\small
\bibliographystyle{IEEEbib}
\bibliography{refs}

\end{document}